\documentclass[aps,pra,twocolumn,nofootinbib]{revtex4-1}
\usepackage{epsfig}
\usepackage[colorlinks,linkcolor=blue,anchorcolor=blue,citecolor=blue,urlcolor=blue,breaklinks=true]{hyperref}
\usepackage{graphics}
\usepackage{color}
\usepackage{amsmath}
\usepackage{bm}
\usepackage{subfigure}
\begin{document}
\author{Guo-Hua Liang$^{1}$}
\email{guohua@nju.edu.cn}
\author{Yong-Long Wang$^{2}$}
\author{Meng-Yun Lai$^{3}$}
\author{Hao Zhao$^{1}$}
\author{Hong-Shi Zong$^{1,4,5}$}
\email{zonghs@nju.edu.cn}
\author{Hui Liu$^{1}$}
\email{liuhui@nju.edu.cn}

\address{$^{1}$ National Laboratory of Solid State Microstructures, School of physics, Collaborative Innovation Center of Advanced Microstructures, Nanjing University, Nanjing 210093, China}
\address{$^{2}$ School of Physics and Electronic Engineering, Linyi University, Linyi 276005, China}
\address{$^{3}$ College of Physics and Communication Electronics Jiangxi Normal University, Nanchang 330022, China}
\address{$^{4}$ Department of Physics, Anhui Normal University, Wuhu, Anhui 241000, China}
\address{$^{5}$ Nanjing Proton Source Research and Design Center, Nanjing 210093, China}

\title{Effective dynamics for a spin-1/2 particle constrained to a space curve in an electric and magnetic field}
\begin{abstract}
We consider the dynamics of a spin-1/2 particle constrained to move in an arbitrary space curve with an external electric and magnetic field applied. With the aid of gauge theory, we successfully decouple the tangential and normal dynamics and derive the effective Hamiltonian. A new type of quantum potential called SU(2) Zeeman interaction appears, which is induced by the electric field and couples spin and intrinsic orbital angular momentum. Based on the Hamiltonian, we discuss the spin precession for zero intrinsic orbital angular momentum case and the energy splitting caused by the SU(2) Zeeman interaction for a helix as examples, showing the combined effect of geometry and external field. The new interaction may bring new approaches to manipulate quantum states in spintronics.
\end{abstract}


\maketitle
\section{INTRODUCTION}
The quantum dynamics in curved spacetime interests scientists for a long time. Based on general relativity, the larger the curvature of spacetime, the stronger the gravitational field. In addition to astronomical surveys, it seems unlikely that we can investigate large curvature effect in conventional laboratories.
While during the last decades, the technique in synthesis of nanostructures has made great progress~\cite{201000591,ph500144s,C4NR00330F,201705630}, which brings  large space curvature to the lab. These nanostructures with curved geometries provide a platform for studies on the dynamics in low-dimensional curved spaces, involving condensed matter~\cite{PhysRevLett.113.227205,PhysRevLett.115.256801,nl504305s}, optics~\cite{PhysRevX.4.011038,Bekenstein2017Control} and magnetism~\cite{PhysRevLett.112.257203,0022-3727-49-36-363001}. Because the curvature radius of the structures may reach nanoscale, nontrivial curvature effects on quantum motion show up, which are not only important in theory aspect, but also indicate great application potential. To show the geometric effects effectively by theory, one successful theoretical approach called the thin-layer procedure (TLP) or confining potential approach for investigating the quantum mechanical properties of particles constrained to low-dimensional curved space was introduced~\cite{JENSEN1971586,PhysRevA.23.1982}.

 TLP originally considers the limiting case of quantum mechanics that a particle in three-dimensional (3D) Euclidean space is subject to a strong confining force acting in the normal direction of a curved surface and gives the effective two-dimensional (2D) Schr\"{o}dinger equation. Interestingly, a geometric potential depending on the intrinsic and extrinsic curvature of the surface appears in this effective equation, showing the geometric effect in constrained systems. Later this potential was demonstrated in photonic crystals~\cite{PhysRevLett.104.150403}. Since TLP was introduced, many theoretical works try to develop and generalize this approach for the application in more situations, such as Schr\"{o}dinger particle~\cite{10.1143/ptp/87.3.561,PhysRevA.89.033630,Wang2018a}, charged particles in an electric and magnetic field~\cite{PhysRevLett.100.230403,Brandt_2015,Wang2016a}, Dirac particles~\cite{OUYANG1999297,PhysRevA.48.1861,BRANDT20163036},spin-1/2 particles with the spin-orbital interaction~\cite{PhysRevB.91.245412,PhysRevB.64.085330,PhysRevB.87.174413,PhysRevA.90.042117,Wang2017,PhysRevA.98.062112} , electromagnetic field~\cite{PhysRevA.78.043821,PhysRevA.97.033843,PhysRevA.100.033825} constrained to space curves and curved surfaces. More general cases of an arbitrary $m$-dimensional manifold embedded in a $n$-dimensional Euclidean space for spinless particles have also been carried out~\cite{S0217732393000891,Maraner_1995,MARANER1996325,S0217751X97002814,SCHUSTER2003132}. It is found that induced SO($n-m$) gauge fields are expected if the normal states are degenerate. Therefore, compared with 2D case, 1D effective dynamics obtained by TLP from 3D Euclidean space shows nontrivial SO(2) gauge potential as an augmented effect.

 The geometrically induced gauge field for 1D optical or electronic waveguides is usually offered by the torsion when only the scalar property is considered, coupling to the intrinsic orbital angular momentum (IOAM)~\cite{PhysRevLett.88.053601,bliokh2015spin,PhysRevLett.118.083601} or topological charge~\cite{PhysRevA.56.4064,PhysRevLett.87.023902,PhysRevLett.111.026802}. While for the motion of spin-1/2 particles in curvilinear coordinates, spin connection acting as a non-Abelian gauge field appears, which is generated by local Lorentz transformation~\cite{16511}. Besides the two geometrically induced gauge fields, magnetic field and spin-orbit interaction due to electric field can also be applied to 1D curved systems in terms of U(1) and non-Abelian gauge fields~\cite{PhysRevB.73.153306,0305,LEURS2008907,0295,RevModPhys.65.733}, respectively. Considering all these effects, the dynamics of a spin-1/2 particle constrained to a space curve in the presence of an electric and magnetic field seems intricate. It is our main aim here to give an effective Hamiltonian for full description of this situation.
 Decoupling from the normal dynamics is the key step in TLP, which does not go well in some cases, especially when external fields are applied. People used to think that the dynamics for a charged particle constrained to a space curve in an electric and magnetic field is difficult to decouple.
 However, it is found in Ref.~\cite{Brandt_2015} that by adopting an appropriate U(1) gauge, one can still get the effective 1D Hamiltonian successfully.
In this paper, we go further to get the effective Hamiltonian for a spin-1/2 particle constrained to a space curve in the presence of an electric and magnetic field, by applying suitable U(1) and SU(2) gauge. We show that the IOAM in a space wire couples to both the magnetic field and the SU(2) gauge field, which induces two types of Zeeman coupling.

The organization of the paper is as follows: In the next section, we derive the effective Hamiltonian for a spin-1/2 particle constrained to a space curve in an electric and magnetic field. In section 3, the spin orientation evolution is calculated for ground state in normal directions based on the effective Hamiltonian. In section 4, the energy band splitting and eigenstates in a helix are discussed. The final section contains a summary.

\section{Effective Hamiltonian for a spin-1/2 particle constrained to a curve}
In this section, we follow the TLP to derive the effective Hamiltonian for a spin-1/2 particle confined to an arbitrary space curve $\mathcal{C}$, including the effect of external magnetic and electric field. The effective Hamiltonian shall be valid for describing dynamics of various 1D semiconducting nanostructures with an electric and magnetic field applied. To make it clear, the analytical derivations for the case without and with an electric and magnetic field are given in turn.
\subsection{Without external fields}
In 3D Euclidean space, the embedded curve $\mathcal{C}$ is parameterized by $\textbf{r}(s)$ with $s$ its arclength. We introduce orthogonal curvilinear coordinates $(s,q_2,q_3)$ and Frenet frame, then the neighbourhood around the curve is described as
\begin{equation}
\bm{R}(s,q_2,q_3)=\textbf{r}(s)+q_2 \textbf{n}(s)+q_3\textbf{b}(s),
\end{equation}
where $\textbf{n}$ and $\textbf{b}$ are the unit normal vector and binormal vector of $\mathcal{C}$, respectively.
Applying the Frenet-Serret equations, we may write
\begin{equation}\label{fren}
\left(
\begin{array}{ccc}
\dot{\textbf{t}}\\
\dot{\textbf{n}}\\
\dot{\textbf{b}}
\end{array}
\right)=\left(
\begin{array}{ccc}
0&\kappa(s)&0\\
-\kappa(s)&0&\tau(s)\\
0&-\tau(s)&0
\end{array}
\right)\left(
\begin{array}{ccc}
\textbf{t}\\
\textbf{n}\\
\textbf{b}
\end{array}
\right),
\end{equation}
where $\textbf{t}$ is the unit tangent vector of $\textbf{r}(s)$, the dot denotes derivative with respect to the natural parameter $s$, and $\kappa(s)$ and $\tau(s)$ are the curvature and torsion of $\mathcal{C}$, respectively.
In this frame, the metric tensor $G_{ij}=\partial_i \bm{R} \cdot \partial_j \bm{R}$ with $i,j=1,2,3$, explicitly reads
\begin{equation}\label{metric}
G_{ij}=\left(
\begin{array}{ccc}
(1-\kappa q_2)^2+\tau^2 (q_2^2+q_3^2) & -\tau q_3 & \tau q_2 \\
-\tau q_3 & 1 & 0 \\
\tau q_2 & 0 &1
\end{array}
\right).
\end{equation}
At each point of the neighbourhood, we can define dreibeins $e_i^I$ corresponding to the the metric tensor $G_{ij}=e_i^I e_j^J \delta_{IJ}$, where $\delta_{IJ}$ is the flat metric and the capital letters $I,J$ denote flat-space indices.

We choose $I,J$ as the locally flat tangent space indices, then the dreibeins for the Frenet frame are written as
\begin{equation}
e_i^I=\left(
\begin{array}{ccc}
1-\kappa q_2 & -q_3\tau & q_2\tau\\
0 & 1 & 0 \\
 0 &0 & 1
\end{array}
\right).
\end{equation}
Inversely, we have
\begin{equation}
e_I^i=\left(
\begin{array}{ccc}
\frac{1}{1-\kappa q_2} & \frac{q_3 \tau}{1-\kappa q_2}  & \frac{-q_2 \tau}{1-\kappa q_2} \\
0 & 1 & 0 \\
 0 &0 & 1
\end{array}
\right).
\end{equation}

In 3D curvilinear coordinates, the non-relativistic equation for a spin-1/2 particle in a confining potential $V_c (q_2,q_3)$ with the contribution of spin connection is
\begin{equation}\label{nonequ}
H=-\frac{1}{2m}[\frac{1}{\sqrt{G}}D_i(\sqrt{G}G^{ij} D_j)]+V_c(q_2,q_3),
\end{equation}
where $D_i=\partial_i+\Omega_i$, with the connection $\Omega_i=\frac{i}{4}\omega_{iIJ} \epsilon^{IJK}\sigma_K$, $G=\det(G_{ij})$. We works in units where $\hbar$ and light speed $c$ are equal to unity throughout the paper. The spin connection
\begin{equation}
\omega_{i IJ}=e_I^j (\partial_i e_{jJ}-\Gamma_{ij}^k e_{kJ}),
\end{equation}
where $\Gamma_{ij}^{k}$ are the usual Christoffel symbols.
The wavefunction of the system $\Phi(s,q_2,q_3)$ should be normalized as
\begin{equation}
\int \sqrt{G} |\Phi|^2 ds dq^2 dq^3=1.
\end{equation}

We assume the confining potential $V_c(q_2,q_3)$ has a deep minimum on $\mathcal{C}$ in the Hamiltonian~\eqref{nonequ}. Next, in the spirit of TLP, we do the rescaling $q_a\rightarrow \sqrt{\epsilon} q_a\ \ (a=2,3)$ and $H\rightarrow G^{1/4}HG^{-1/4}$, and introduce the new wave function $\Psi = G^{1/4}\Phi$. The parameter $\epsilon$ is assumed sufficiently small, and the confining potential $V_c(q_a)\gg \frac{1}{2m q_a^2\epsilon}$.

Before performing TLP, we have to give the explicit form of the spin connection. After some straightforward calculations, we find
\begin{equation}\label{spinc}
\begin{aligned}
\Omega_s=\frac{i}{2}(-\kappa\sigma_{\bar{3}}-\tau \sigma_{\bar{s}})+O(\epsilon^{1/2}), \ \
\Omega_2=\Omega_3=0.
\end{aligned}
\end{equation}
To be clear, the subscripts with a bar stand for the local flat indices $I,J$.
It turns out that only the tangential component of the connection is nonzero, which makes the TLP easy to perform in this case.

Substituting the metric~\eqref{metric} and Eqs.~\eqref{spinc} into the Hamiltonian \eqref{nonequ} and expanding the rescaled Hamiltonian in the order of $\epsilon$, we obtain
\begin{equation}
H=\frac{1}{\epsilon}H^{(-1)}+H^{(0)}+O(\epsilon^{1/2}),
\end{equation}
where
\begin{equation}\label{heff0}
H^{(-1)}=-\frac{1}{2m}(\partial_2^2+\partial_3^2)+\epsilon V_c(q_2,q_3),
\end{equation}
and
\begin{equation}\label{Heff1}
H^{(0)}=-\frac{1}{2m}(\partial_s+\Omega_s-i\tau \hat{L})^2+V_g,
\end{equation}
wherein the angular momentum operator is defined as $\hat{L}=-i (q_2\partial_3-q_3 \partial_2)$, and the geometric potential $V_g=-\frac{\kappa^2}{8m}$. $H^{(-1)}/\epsilon$ is the 2D Hamiltonian for a particle confined by the potential $V_c$, while Eq.~\eqref{Heff1} describes the dynamics of a spin-1/2 particle bounded to the curve.

Eq.~\eqref{Heff1} is still not the effective 1D Hamiltonian required since the operator $\hat{L}$ is composed of normal derivatives. To get the availably effective Hamiltonian, we have to consider the eigenstates of the normal Hamiltonian~\eqref{heff0}.
Because $V_c$ is independent of $s$, we separate the rescaled wavefunction into tangential and normal wavefunctions, that is
\begin{equation}\label{psis}
\Psi=\sum_\beta \psi_\beta(s)\chi_\beta (q_2,q_3),
\end{equation}
where the index $\beta$ labels the degeneracy in the spectrum of the normal Hamiltonian $H^{(-1)}$. The wavefunction $\chi(q_2,q_3)$ is totally determined by the confinement $V_c$. Here we only consider the case of a circular cross section, that is $V_c$ with SO(2) symmetry. More general case of the cross section has been discussed in Ref.~\cite{10.1143/ptp/87.3.561} for Schr\"{o}dinger equation. In this case it is convenient to make a coordinate transformation in the normal plane $(q_2,q_3)\rightarrow (r,\varphi) $, where $r=\sqrt{(q_2)^2+(q_3)^2}, \varphi=\arctan (q_2/q_3)$. Then the normal eigenstates can be written as $\chi(r,\varphi)_{nl} =\frac{1}{\sqrt{2\pi}}R_n(r)e^{il\varphi}$, where $R_n(r)$ are normalized radial wavefunctions, $n$ and $l$ are radial and angular quantum numbers, respectively. In the Hilbert space spanned by these normal eigenstates, the Hamiltonian~\eqref{Heff1} becomes a matrix with the elements
\begin{equation}
H^{(0)}_{nln^\prime l^\prime}=\int drd\varphi r \chi_{nl} H^{(0)} \chi_{n^\prime l^\prime}.
\end{equation}
It is easy to find this matrix is diagonal, so we can write the effective Hamiltonian as
\begin{equation}\label{heff2}
H_\text{eff}=-\frac{1}{2m}[(\partial_s+\Omega_s+i\tau l )^2+\frac{\kappa^2}{4}].
\end{equation}
From Eq.~\eqref{heff2}, we find that for a spin-1/2 particle constrained to a space curve with a circular cross section, two geometrically induced gauge fields appear in the effective Hamiltonian, which can be separated into spin angular momentum (SAM) part and IOAM part. The SAM part depends on both the curvature and torsion of the space curve, while the IOAM part depends only on the torsion.

\subsection{With an electric and magnetic field}
In this section, we consider the case with an external electric field $\bm{E}$ and magnetic field $\bm{B}$ applied. By introducing the SU(2) gauge field~\cite{frohlich1992u,RevModPhys.65.733,LEURS2008907}, the non-relativistic Hamiltonian for a spin-1/2 particle in an electric and magnetic field, is of the form
\begin{equation}\label{hem}
\begin{aligned}
H=&-\frac{1}{2m}\frac{1}{\sqrt{G}}D_i(\sqrt{G}G^{ij}D_j)-\mu_B \textbf{B}\cdot \bm{\sigma} +eV,
\end{aligned}
\end{equation}
where $e$ is electric charge, $\mu_B$ is the Bohr magneton, the covariant derivative $D_i=\partial_i+\Omega_i-ieA_i+i\frac{e}{4m}\epsilon_{ijk}\sigma^j E^k$, wherein $A_i$ is the magnetic vector potential, and the last gauge term accounts for the spin-orbit interaction from electric field. The second term in Eq.~\eqref{hem} is the usual Zeeman coupling, and $V$ is the scalar potential. Here we neglect the Darwin term and higher order corrections.
Formally the gauge field can be divided into two parts, one is the $U(1)$ gauge field $A_i$, the other is the SU(2) gauge field $\Omega_i+i\frac{e}{4m}\epsilon_{ijk}\sigma^j E^k$.
We denote that $W_i=i\Omega_i-\frac{e}{4m}\epsilon_{ijk}\sigma^j E^k$.
In Yang-Mills gauge field theory, the covariant derivative is written as $D_\mu=\partial_\mu-igA_\mu^i \frac{\sigma^i}{2}$, with $g$ the coupling constant. Following Yang-Mills theory, we can write the tangential and normal components of gauge field as $W_s=i\Omega_s-\lambda w_{sj}\frac{\sigma^j}{2}$ and $W_a=-\lambda w_{aj}\frac{\sigma^j}{2}$, where $\lambda=-\frac{e}{2m}$ and $w_{ij}=\epsilon_{ijk}E^k$.

Before performing the confining potential approach, we have to note the gauge freedom of electromagnetic vector potentials. Thus we need to expand the electromagnetic field potential in the vicinity of the space curve, that is
\begin{equation}
A_i(s,\sqrt{\epsilon} q_a)=A_i(s,0)+\sqrt{\epsilon} q_a \partial_a A_i(s,q_b)|_{q_b=0}+O(\epsilon).
\end{equation}
Similarly, for the SU(2) gauge field, we can also expand it as
\begin{equation}
W_i (s,\sqrt{\epsilon} q_a)= W_i(s,0)+\sqrt{\epsilon} q_a \partial_a W_i(s,q_b)_{q_b=0}+O(\epsilon).
\end{equation}

Again, we introduce the confining potential $V_c$ and expand the rescaled Hamiltonian up to zeroth order of $\epsilon$ and obtain
\begin{equation}\label{hexp}
H=\frac{1}{\epsilon}H^{(-1)}+\frac{1}{\sqrt{\epsilon}}H^{(-1/2)}+H^{(0)}+O(\epsilon^{1/2}),
\end{equation}
where
\begin{equation}
H^{(-1)}=-\frac{1}{2m}(\partial_2^2+\partial_3^2)+\epsilon V_c(q_2,q_3),
\end{equation}
\begin{equation}\label{12od}
H^{(-1/2)}=\frac{i}{m}(eA_a+W_a)\partial_a,
\end{equation}
and
\begin{equation}\label{tham}
\begin{aligned}
H^{(0)}=&-\frac{1}{2m}[(D_s+i\tau \hat{L})^2+ \frac{\kappa^2}{4} ]+eV\\
&-\frac{1}{2m}[(-ieA_a-iW_a)^2 +\partial_a(-ieA_a-iW_a)] \\
&+\frac{i}{m}q_b\partial_b (e A_a+W_a)\partial_a -\mu_B \textbf{B}\cdot \bm{\sigma}.
\end{aligned}
\end{equation}
Comparing with the case without external field, we find a term in the order of $\epsilon^{-1/2}$ appears in the expression due to the external field applied.
It seems from Eq.~\eqref{12od} that this term and the terms containing derivatives respect to normal coordinates  in Eq.~\eqref{tham} prevent the separation between the tangent and normal dynamics.
In the following we seek appropriate gauge for successful separation of the dynamics.

For the U(1) gauge field, we can find a gauge transformation $A^\prime_i=A_i+\partial_i \gamma$, $\psi^\prime=\psi e^{ie\gamma}$ where
\begin{equation}
\gamma=-A_a \sqrt{\epsilon}q^a+\frac{1}{2}\epsilon q^aq^b \partial_aA_b.
\end{equation}
Then after the gauge transformation the electromagnetic field becomes
\begin{equation}\label{aps}
A^\prime_s=A_s+O(\epsilon^{1/2}),
\end{equation}
\begin{equation}\label{aapr}
A_a^\prime=-\sqrt{\epsilon}\frac{q^b}{2} F_{ab}+O(\epsilon),
\end{equation}
where $F_{ab}=\partial_aA_b-\partial_b A_a$ is the electromagnetic field tensor.

Next we focus on the SU(2) gauge field $W_a$. Corresponding to the infinitesimal form of the fermion transformation $\psi\rightarrow (1+\alpha_i \frac{\sigma^i}{2})\psi$, the transformation of the gauge field should be $w_{ai}\frac{\sigma^i}{2} \rightarrow w_{ai}\frac{\sigma^i}{2}+\frac{1}{\lambda}(\partial_a \alpha_i \frac{\sigma^i}{2})+i[\alpha_i \frac{\sigma^i}{2}, w_{aj}\frac{\sigma^j}{2}]$. Now we define the SU(2) gauge transformation
\begin{equation}
\begin{aligned}
\alpha_i\frac{\sigma^i}{2}=&-\lambda \sqrt{\epsilon}q^a w_{ai}\frac{\sigma^i}{2}+ \frac{\lambda}{2}\epsilon q^aq^b\partial_a(w _{bi}\frac{\sigma^i}{2}).
\end{aligned}
\end{equation}

Applying this gauge transformation, we find the gauge field becomes
\begin{equation}\label{wap}
W^\prime_a=-\frac{1}{2}\sqrt{\epsilon}q^b F_{abi}\frac{\sigma^i}{2},
\end{equation}
where we define the SU(2) field strength as
\begin{equation}
F_{abi}\frac{\sigma^i}{2}=\partial_a (w_{bi}\frac{\sigma^i}{2})-\partial_b (w_{ai}\frac{\sigma^i}{2})-2i\lambda [w_{aj} \frac{\sigma^j}{2},w_{bk}\frac{\sigma^k}{2}].
\end{equation}
As in the case of Eq.~\eqref{aps}, for the tangential component $W_s$, we can also obtain $W_s^\prime=W_s+O(\epsilon^{1/2})$ after the corresponding transformation. Hence after the gauge transformations we adopt, the tangential components of the gauge field remain unchanged.


Substituting Eq.~\eqref{aapr} and Eq.~\eqref{wap} into the expansion~\eqref{hexp}, it is found the $\epsilon^{-1/2}$ order term vanishes and
\begin{equation}\label{tham2}
\begin{aligned}
H^{(0)}=&-\frac{1}{2m}[(D_s+i\tau \hat{L})^2+ \frac{\kappa^2}{4}]+eV \\
&+\lambda(F^{ab}L_{ab}+F^{ab}_i\frac{\sigma^i}{2} L_{ab} )-\mu_B \textbf{B}\cdot \bm{\sigma},
\end{aligned}
\end{equation}
where $L_{ab}=-i (q_a \partial_b-q_b \partial_a)$.

To obtain the effective Hamiltonian for the tangential dynamics, we still do the same procedure as the process from Eq.~\eqref{psis} to Eq.~\eqref{heff2}.
Note that the field strength $F^{ab}$ and $F^{ab}_i\frac{\sigma^i}{2}$ can be explicitly expressed into expressions of the external electric and magnetic field. The final form of the effective Hamiltonian for spin-1/2 particles constrained to a space curve in the presence of an electric and magnetic field is
\begin{equation}\label{heff3}
H_{\text{eff}}=-\frac{1}{2m}[(D_s+i\tau l)^2+ \frac{\kappa^2}{4}]+eV+H_z,
\end{equation}
where
\begin{equation}~\label{hz}
H_z=-\mu_B \textbf{B}\cdot \bm{\sigma}+2\lambda B_s l+2\lambda F_{so}l,
\end{equation}
and
\begin{equation}
F_{so}=(\bm{\nabla}_\perp \cdot \bm{E}_\perp)\frac{\sigma_s}{2}-(\frac{\bm{\sigma}_\perp}{2}\cdot \bm{\nabla}_\perp)E_s+\lambda (\bm{E}\cdot \bm{\sigma})E_s,
\end{equation}
wherein $\perp$ stands for coordinates $(q_2,q_3)$ in the normal plane of the curve, and $B_s=\bm{B}\cdot \bm{t}$ is the tangential component of the magnetic field. We can find that $H_z$ contains three parts. The first one is the usual Zeeman coupling term composed of the magnetic field and spin angular momentum. The second one is an induced Zeeman interaction between tangential magnetic field and IOAM. The third one is a new type of Zeeman interaction discovered in this paper, which is a coupling between IOAM and the SU(2) field strength, and we refer to as SU(2) Zeeman interaction. The necessary conditions of this interaction are $l\neq 0$ and the non-zero gradient of the electric field. Because of the new Zeeman interaction, the spin and intrinsic orbital angular momentum do not evolve independently anymore. Eq.~\eqref{heff3} is the key result of the present paper, and the following discussions are based on this effective Hamiltonian.


\section{spin precession for $l=0$}
In this section, we study the spatial behavior of spin precession for moving particles constrained to a space curve in the case of $l=0$. To distinguish the effects of geometry and external electric field, we still consider the case without and with electric field in turn. No magnetic field is applied in this section.
\subsection{Without external fields}
Without external fields, the effective Hamiltonian~\eqref{heff2} differs from the 1D free-electron Hamiltonian $H_f=-\frac{1}{2m}\partial_s^2$ only by a spin connection $\Omega_s$ as a gauge field and a geometric potential $V_g$. Therefore, by constructing a unitary transformation operator $U=e^{-\int  \Omega_s ds }$,
we obtain $U^\dagger H_{\text{eff}} U=H_f+V_g$. Correspondingly, if $\psi$ is the eigen wave function of $H_{\text{eff}}$,  $\psi_f=U^\dagger \psi$ is the eigenstate of $H_f+V_g$, retaining the spin state. This means the precession of the electrons can be described by the unitary transformation $U$. Considering the explicit form of $\Omega_s$, we can always write the unitary transformation as
\begin{equation}
U=e^{\frac{i\bm{\sigma}\cdot\bm{h}\phi}{2}},
\end{equation}
where $\phi=\sqrt{\phi_c^2+\phi_t^2}$ with $\phi_c=\int \kappa ds $ and $\phi_t=\int \tau ds$, and $\bm{h}=(-\tau/\phi,0,-\kappa/\phi)^T$ is a unit vector in the Frenet coordinates.
By using the formula
\begin{equation}
\exp(\frac{i\bm{\sigma}\cdot\bm{h}\phi}{2})=\cos(\phi/2)+i\bm{\sigma}\cdot \bm{h}\sin(\phi/2),
\end{equation}
we can directly calculate the spin orientation
\begin{equation}
\langle \bm{\sigma} \rangle =\langle \psi_f |U^\dagger \bm{\sigma} U| \psi_f \rangle.
\end{equation}
Further, the spatial derivative of the expectation value of spin components are obtained as
\begin{equation}
\partial_s \langle \bm{\sigma} \rangle = \langle [\Omega_s, \bm{\sigma}]\rangle +\langle \partial_s \bm{\sigma}\rangle.
\end{equation}
We emphasize here that, the spin connection is dreibein dependent.
If one chooses the local flat tangent space coordinates, the spin connection is the form in Eq.~\eqref{spinc}, and $\partial_s \bm{\sigma}=(\partial_s e_I^i)\sigma^I =O(\epsilon^{1/2})$. Therefore, the commutator $\langle [\Omega_s, \bm{\sigma}]\rangle$ accounts for the precession. One can also choose the frame where the spin connection vanishes~\cite{PhysRevB.94.081406}, and the final results are equivalent.
Hence the spatial derivative of the spin orientation expectation has a matrix form
\begin{equation}\label{roaxm}
\left(
\begin{array}{ccc}
\partial_s \langle \sigma_s \rangle \\ \partial_s \langle \sigma_2 \rangle \\ \partial_s \langle \sigma_3 \rangle
\end{array}
\right)=\left(
\begin{array}{ccc}
0 & \kappa & 0 \\ -\kappa & 0 &\tau \\ 0 & -\tau & 0
\end{array}
\right) \left(
\begin{array}{ccc}
\langle \sigma_s \rangle \\ \langle \sigma_2 \rangle \\ \langle \sigma_3 \rangle
\end{array}
\right)
\end{equation}
or a compact form
\begin{equation}\label{roax}
\partial_s \langle \bm{\sigma}\rangle=\phi\bm{h}\times \langle \bm{\sigma}\rangle.
\end{equation}
From Eq.~\eqref{roax} we can see that $\bm{h}$ is in fact the instantaneous axis of rotation for spin orientation. Comparing Eq.~\eqref{roaxm} with Eq.~\eqref{fren}, we find that they have a similar form. However, it should be note that in Eq.~\eqref{fren} the elements on the left side are derivatives of vectors, and in Eq.~\eqref{roaxm} they are derivatives of spin orientation expectation components. Therefore the rotation of spin orientation may be different from the rotation of the vector in Frenet frame.

\begin{figure}
  \centering
  \includegraphics[width=0.45\textwidth]{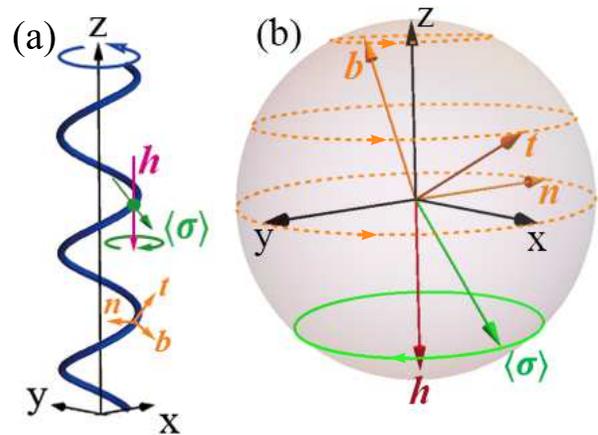}\\
  \caption{(a) Schematic diagram of a helix. The parameter values are $r_0=1 $, $d=0.5$. Here the length unit is arbitrary. (b) Evolution of the spin orientation on the Bloch sphere for a spin-1/2 particle constrained to the helix .}\label{fig1}
\end{figure}

To make it clear, we assume the space curve $\mathcal{C}$ is a helix (see Fig.~\ref{fig1}(a)), which can be described in Cartesian coordinates as
\begin{equation}
x=r_0 \cos\theta \ \ y=r_0 \sin\theta \ \  z=d \theta.
\end{equation}
It is easy to obtain the curvature $\kappa=\frac{r_0}{r_0^2+d^2}$, the torsion $\tau=\frac{d}{r_0^2+d^2}$, and the arclength $s=\sqrt{r_0^2+d^2}\theta$. We would like to exhibit the variation of the expectation value of the spin orientation in Cartesian coordinates, which would give a intuitional picture in a laboratory frame. In Cartesian coordinates, the spin gauge potential is simply written as $\Omega_s=-\frac{i}{2\sqrt{r_0^2+d^2}}\sigma_z$, since the axis of rotation $\bm{h}$ is found to be $-\bm{e}_z$. Then we get the evolution of the spin orientation for a moving spin-1/2 particle constrained to a helix, which is shown in Fig.~\ref{fig1}(b). It is found that moving along a helix leads to a spin orientation rotation in the opposite direction respect to the rotation of the helix frame.

\subsection{With a radial electric field}
\begin{figure}
  \centering
  \includegraphics[width=0.45\textwidth]{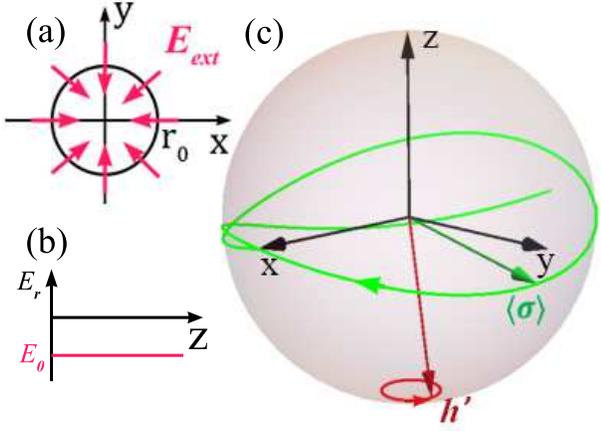}\\
  \caption{(a) Schematic picture of the applied radial electric field in x-y plane. The parameter values are $r_0= l_E $, $d=0.5l_E$ with the length unit $l_E=1/2\lambda E_0$. (b)The independence of the electric field on $z$.  (c)Evolution of spin orientation $\langle \bm{\sigma}\rangle$ and axis of rotation $\bm{h}^\prime$ on the Bloch sphere in the lab frame.}\label{fig2}
\end{figure}

Now, for simplicity, we apply a uniform radial electric field $\bm{E}_r$ to the helix (see Fig.~\ref{fig2}(a)). In this case, $W_s =i\Omega_s+ \frac{\lambda}{2} \sigma_3E_{0}$ with $E_0$ the intensity of $\bm{E_r}$ at $r=r_0$. Since we assume the applied radial electric field is inward in the x-y plane and uniform along $z$ direction, $E_0$ is negative (see Fig.~\ref{fig2}(b), the positive direction is outward). The corresponding unitary transformation in this case should be $U^\prime=e^{i\int W_s ds}$. We write the unitary transformation as $U^\prime=e^{\frac{i\bm{\sigma}\cdot\bm{h}^\prime\phi^\prime}{2}}$, with
\begin{equation}
\bm{h}^\prime=(\frac{-\tau}{\phi^\prime},0,\frac{-\kappa-\lambda E_0}{\phi^\prime})^T,
\end{equation}
and $\phi^\prime=\sqrt{\phi_c^{\prime 2}+\phi_t^2}$, where $\phi_c^\prime=\int \kappa+\lambda E_0 ds$. The vector $\bm{h}^\prime$ is the instantaneous axis of rotation in the case that a radial electric field is applied. The variation of spin orientation in this case is shown in Fig.~\ref{fig2}(c) for the length range $0<\theta <2\pi$, with the initial orientation along $x$ direction. One readily notes that the axis of rotation is no longer fixed but rotates with the particle moving. Because of the rotation of axis $\bm{h}^\prime$, the orientation of spin evolves more complex than the case without external field. In addition, another difference caused by the external electric field is the extra rotation phase in $\phi_c^\prime$. This is why we find that the spin return to its original orientation at a certain position $\theta_0$ with $\theta_0< 2\pi$, showing the property of Rashba spin-orbit interaction. The spin precession in the case indicates that the curvature is tantamount to provide an effective electric field.

\section{Energy splitting and eigen states}
In this section we consider the effective dynamics in a helix for $l=\pm{1}$ and mainly pay attention to the induced SU(2) Zeeman interaction. To ensure such a coupling term plays a role, we apply the helix a radial electric field whose radial gradient is nonzero.
In the following, we do a perturbation calculation on the energy levels 
in the helix for $l=\pm{1}$.

The unperturbed Hamiltonian is $H_0=-\frac{1}{2m}(D_s+i\tau l)^2-\frac{ \kappa^2}{8m}+eV$. We assume that the induced SU(2) Zeeman coupling and the magnetic field are sufficiently weak that they can be treated as a perturbation. In this case, according to Eq.~\eqref{heff3}, the perturbation is written as $H^\prime=\lambda (\partial_r E_r)_{r=r_0}\sigma_s l+2\lambda B_s l-\mu_B \bm{B}\cdot \bm{\sigma}$.

\begin{figure}
  \centering
  \includegraphics[width=0.46\textwidth]{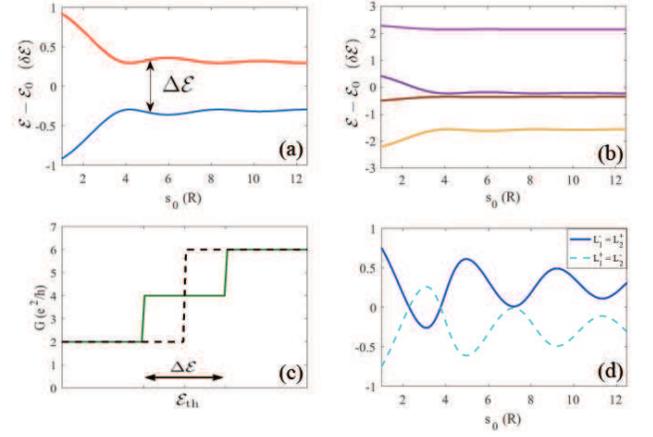}\\
  \caption{Energy splitting in a helix as a function of total length for $l=\pm1$ with (a) only electric field and (b) both the electric and magnetic field ($\mu_B B_z=\delta \mathcal{E}$) applied. The unit $\delta \mathcal{E}=|\lambda (\partial_r E_r)|_{r=r_0}$ and $R=\sqrt{r_0^2+d^2}$. The parameter values are $r_0= l_E $, $d=0.5l_E$. (c) Ballistic conductance at the vicinity of the threshold energy $\mathcal{E}_{th}$, in the case of $\partial_r E_r=0$ (dash line) and $\partial_r E_r\neq0$ (solid line) gradient. (d) The expectation values of intrinsic orbital angular momentum $L_1^-=L_2^+$ (solid line) and $L_1^+=L_2^-$ (dash line) for the basis of eigenvectors.}\label{fig3}
\end{figure}

Firstly, we assume $\bm{B}=0$. Since $\kappa$ and $V$ are constant for the helix, it is easy to solve the eigen equation $H_0 |s_t,l\rangle =\mathcal{E}_0 |s_t,l\rangle $, and find the eigenvalue $\mathcal{E}_0=\frac{1}{2m} (k^2-\kappa^2/4)+eV$, and the degenerate eigenstates with the form $|s_t ,l\rangle=e^{iks}e^{il\varphi}e^{i\int W_s ds}|s_t \rangle$, where $s_t =+, -$ with the definition $|+\rangle=(1,0)^T $ and $|- \rangle =(0,1)^T$. From degenerate perturbation theory, we obtain
\begin{equation}\label{purt}
\sum_{s_t^\prime, l^\prime} [H^\prime_{s_t^\prime l^\prime,s_t l}-(\mathcal{E}-\mathcal{E}_0)\delta_{s_t^\prime l^\prime,s_t l}]a_{s_t l}=0,
\end{equation}
with $\mathcal{E}$ the eigenvalue of $H_0+H'$ and $a_{s_t l}$ the zeroth-order coefficients used to expand the perturbed states in terms of $|s_t ,l\rangle$. Eq.~\eqref{purt} yields eigenvalues $\mathcal{E}^\pm=\mathcal{E}_0\pm \Delta \mathcal{E}/2$, which is shown in Fig.~\ref{fig3}(a) as a function of the total length $s_0$. It is shown that the energy gap $\Delta \mathcal{E}$ varies with $s_0$ initially, and later tend to a definite value. The initial dependence on the total length is caused by the spin precession we present in the above section. This energy gap can be utilized to demonstrate the SU(2) Zeeman interaction experimentally. When the gradient of the electric field $(\partial_r E_r)$ is zero, with the increasing of total energy, the conductance of the helical wire shows a step-like structure at the threshold energy $\mathcal{E}_{th}$ for $l=\pm1$ (see the dash line in Fig.~\ref{fig3}(c)). While if we increase the value of $(\partial_r E_r)|_{r=r_0}$, a new plateau will appear in the conductance curve (the solid line), and the width of the plateau is $\Delta \mathcal{E}$.

The energy splitting caused by the SU(2) Zeeman coupling reduce the degeneracy of the system from 4 to 2. The remaining degeneracy is due to the property of the SU(2) Zeeman interaction that its actions on $|s_t,l \rangle$ and $|-s_t,-l \rangle$ are the same. The associated eigenvectors are therefore still not determined completely because of the remaining degeneracy. However, we can write the basis of eigenvector for the energy $\mathcal{E}^+$ as
\begin{equation}
|\psi_1^+\rangle=a_o^+|+,-1 \rangle +a_e^+ |+,+1\rangle,
\end{equation}
\begin{equation}
|\psi_2^+\rangle=a_o^+|-,+1 \rangle +a_e^+ |-,-1\rangle,
\end{equation}
and the basis for the energy $\mathcal{E}^-$ as
\begin{equation}
|\psi_1^-\rangle=a_o^-|+,-1 \rangle+a_e^-|+,+1\rangle,
\end{equation}
\begin{equation}
|\psi_2^-\rangle=a_o^-|-,+1 \rangle+a_e^-|-,-1 \rangle,
\end{equation}
where $a_o^\pm$ and $a_e^\pm$ can be determined from Eq.~\eqref{purt}. To show the effect of the SU(2) Zeeman interaction on states, we give the expectation of the intrinsic angular momentum for each eigenstate basis in Fig.~\ref{fig3}(d), namely $L_\mu^\pm=\langle \psi_\mu^\pm|\hat{L}|\psi_\mu^\pm \rangle$ with $\mu=1,2$ the degeneration index. It shows that $L_\mu^\pm$ oscillate with the total length and the amplitude tend to reduce with $s_0$ increasing. Because of the degeneracy, we can find the relation $L_1^\pm=L_2^\mp$.
When the total energy in the range between $\mathcal{E}_{th}-\Delta \mathcal{E}/2$ and $\mathcal{E}_{th}+\Delta \mathcal{E}/2$, the channels $|\psi_1^-\rangle$ and $|\psi_2^-\rangle$ are open. In this case if we drive purely spin-polarized particles into the system, the IOAM polarization may be expected.

The discussion above is for the case without external magnetic field. Unlike the SU(2) Zeeman interaction induced by the radial electric field with non-zero gradient, Zeeman coupling for spin and IOAM due to magnetic field could relieve the degenerate energy levels completely. In Fig.~\ref{fig3}(b) we add the effect of a magnetic field $B_z$, and plot the energy splitting against the total length. Comparing Fig.~\ref{fig3}(a) and (b), it is obvious that the magnetic field relieves the partially degenerate energy levels in Fig.~\ref{fig3}(a) and also break the symmetry of the energy levels about $\mathcal{E}=\mathcal{E}_0$.

\section{Conclusion}
To conclude, we have performed thin-layer procedure to derive the effective Hamiltonian for a spin-1/2 particle constrained to a space curve in the presence of an electric and magnetic field. The difficulty on separation of the dynamics in tangential and normal direction is overcame by a suitable choice of gauges. The final result shows that a new quantum potential induced by the external electric field appears in the effective dynamics, which couples the spin and intrinsic orbital angular momentum, and can be described as the SU(2) Zeeman interaction. Based on the effective Hamiltonian, we have shown the spin precession in a helix without and with a radial electric field applied, in the case of zero intrinsic orbital angular momentum. It shows that the curvature can perform the role of electric field, and the radial electric field rotates the instantaneous axis of rotation for the expectation of spin orientation. For the first excited modes $l=\pm1$, the energy splitting due to the SU(2) Zeeman interaction in a helix have been discussed. The SU(2) Zeeman interaction relieves the degeneracy partially and does not break the time reversal symmetry, showing different effect from the magnetic field. We expect that these finds would provide new routes for manipulating spin, intrinsic orbital angular momentum in nanometer devices.

\acknowledgments

This work is supported in part by the National Natural Science Foundation of China (under Grants No. 11690030, No. 11475085, No. 11535005 and No. 61425018).

\bibliographystyle{apsrev4-1}
\bibliography{reference}

\end{document}